\documentclass{SCIYA2023enOL}
 \online
 \usepackage{lineno}

 \usepackage{multirow}
 \usepackage{hyperref}
 \usepackage{amsmath}
\usepackage{tikz}
\usetikzlibrary{arrows.meta, positioning, shapes, calc}
\usepackage{setspace} 
\hypersetup{colorlinks=true, linkcolor=blue, urlcolor=blue}
\begin{document}

\ensubject{fdsfd} 
\ArticleType{ARTICLES}
\Year{2025}
\Month{January} 
\Vol{65}
\No{1}
\BeginPage{1}  
\DOI{Science China Mathematics}
\title[]{Dynamical Insights into the Current Control Strategies for
Chikungunya}{Dynamical Insights into the Current Control Strategies for
Chikungunya}
\author[1,2]{Bo Zheng}{zhengbo611@outlook.com}
\author[1,4]{Xingzhi Chen}{chenxingzhi.math@gmail.com}
\author[1,2]{Jianglong Xiao}{jlx@e.gzhu.edu.cn} 
\author[1,2]{Kai Wang}{kwang@gzhu.edu.cn}
\author[1]{Jianshe Yu}{jsyu@gzhu.edu.cn}
\author[3,4,$\ast$]{Huaiping Zhu}{huaiping@yorku.ca}
\AuthorCitation{Bo Zheng,  Xingzhi Chen,  Jianglong Xiao,  et al}
\address[1]{Center for Applied Mathematics, Guangzhou University, Guangzhou, {\rm 510006}, China}
\address[2]{School of Mathematics and Information Science, Guangzhou University, Guangzhou, {\rm 510006}, China}
\address[3]{Department of Mathematics and Statistics, York University, Toronto, ON, {\rm M3J 1P3}, Canada}
\address[4]{Laboratory of Mathematical Parallel Systems (LAMPS) and Center of Diseases Modeling (CDM),\\ York University, Toronto, ON, {\rm M3J 1P3}, Canada} 
 

\abstract{This paper addresses the challenge that, under China's current
mosquito-control policy for chikungunya---%
\textit{clearing stagnant water, killing adult mosquitoes, and preventing
bites}---epidemics can still recur and continue to rise. By constructing a
vector--host dynamical model that incorporates asymptomatic
infections and a screening rate, we derive approximate analytical
expressions for the epidemic peak value and peak time, and reveal the
intrinsic link between short-term transients and long-term outcomes.
Simulations across three scenarios---human-seeded, mosquito-seeded, and joint
human--mosquito-seeded---show that (i) even when the vector population is
effectively suppressed, asymptomatic carriers can still sustain transmission;
(ii) the effect of the screening rate on the peak time is non-monotonic, and there exists a critical threshold that maximizes the delay of the epidemic; and (iii) moderate screening can both
suppress the first wave and avoid leaving an excessively large susceptible
pool that would amplify subsequent waves. Accordingly, we upgrade the
original policy to a \textit{Twelve-character} strategy for prevention,
mitigation, and control---\textit{clearing stagnant water, killing adult
mosquitoes, preventing bites, and screening cases early}---providing theoretical support for chikungunya and other mosquito-borne diseases.} 



\maketitle
\section{Introduction}\label{sec1}
In 2025, China experienced its largest recorded outbreak of chikungunya fever (CHIK). From the first reported local case on July 8 to October 18, Guangdong Province reported a cumulative total of 23,976 local cases, mainly concentrated in three cities: Foshan (10,626 cases), Jiangmen (9,739 cases), and Guangzhou (1,118 cases). After a rapid rise, the outbreak in Foshan gradually subsided, with daily new cases falling below 50 after August 17 and remaining at that level, suggesting that local control measures had produced some effect. However, as the outbreak in Foshan faded, the epidemic spread to neighboring areas, and cases in Jiangmen rose again, with 2,480 new cases reported in the single week from September 28 to October 4, highlighting the rapid regional spread of the epidemic. Subsequent reports from the Guangdong Provincial Center for Disease Control and Prevention (CDC) showed that 2,086 local cases were still reported across the province from October 12 to 18, 2025, indicating that transmission had not yet been effectively interrupted \cite{GDCDC}. 

CHIK is an acute infectious disease caused by the chikungunya virus (CHIKV). Similar to dengue fever (DF), CHIK is not transmitted directly from person to person, but through the bites of {\it Aedes} mosquitoes \cite{Cag2013NewM, Constant2021}. In response to the rapid spread of the 2025 outbreak in Guangdong, and to quickly reduce mosquito density and minimize the risk of transmission and spread, the relevant authorities immediately launched a public health campaign focused on eliminating mosquito breeding sites and killing adult mosquitoes. This campaign followed the same strategy adopted during the 2014 dengue outbreak, namely the ``Nine-character'' strategy of {\it clearing stagnant water, killing adult mosquitoes, and preventing bites}. 

 \begin{figure}[htbp]
    \centering
    \includegraphics[height=6cm, width=14cm]{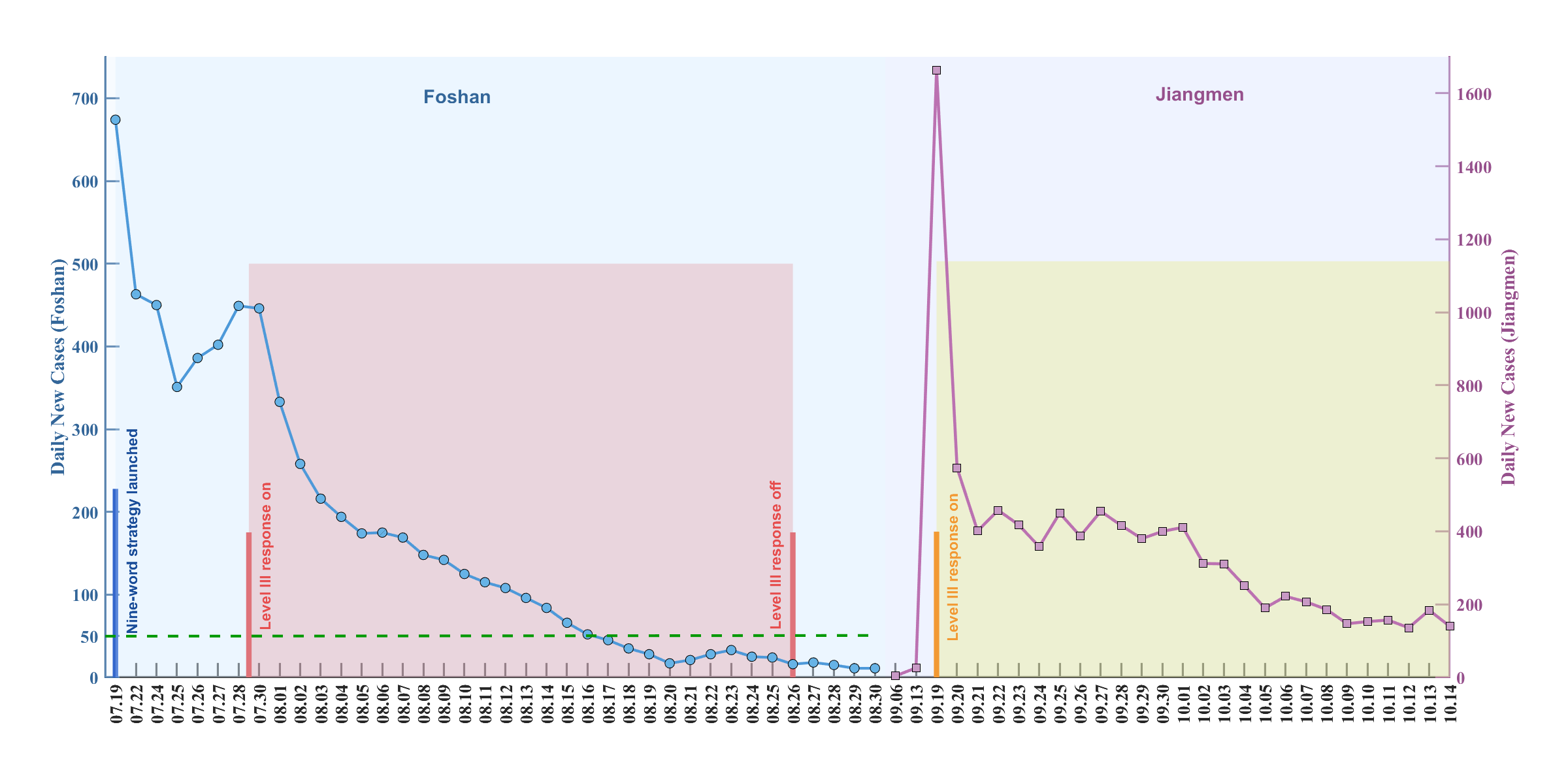} 
\caption{{\bf Daily new cases of chikungunya reported in Foshan and Jiangmen, Guangdong, in 2025.}   Foshan carried out the {\it Nine-character strategy}
 on July 19.  The Level III public health emergency response, which is implemented when a significant public health event has occurred, in Foshan was initiated on July 29 and terminated on August 26, whereas Jiangmen activated its Level III response on September 19.  Data are taken from the Guangdong CDC  \cite{GDCDC}.}
\label{Fig1} 
\end{figure} 

However, after the implementation of this strategy, observations in Foshan from July 19 to 28 showed a rebound in case numbers, suggesting that these interventions did not immediately bring the outbreak under control, as shown in Figure \ref{Fig1}.  This situation is similar to the 2014 dengue outbreak in Guangzhou: the latter also failed to be rapidly contained after control measures were implemented, lasting from June 11 to December 21, a total of 194 days, peaking in October, and only being effectively controlled in early November after daily cases fell below 300 \cite{LinkofXinhua}.  Under this situation, a tough question arises: why does the epidemic continue to rise even under strict enforcement of the ``Nine-character'' strategy? This paper aims to re-examine and improve the traditional control measures represented by the ``Nine-character'' strategy, providing new theoretical evidence and decision-making references for future epidemic prevention and control. 

The {\it Nine-character strategy} aligns with the theoretical findings of Ross–Macdonald in \cite{Ross1899, Smith2012}, which claimed that {\it in order to eliminate malaria wholly or partly from a given locality, it is necessary only to exterminate the various species of insects which carry the infection. } However, it is often overlooked that the Ross–Macdonald model was originally developed to investigate malaria transmission dynamics and has played a crucial role in shaping strategies while substantially advancing our understanding of malaria control and elimination.   The pioneering work of Ross–Macdonald established the mathematical foundations of malaria transmission dynamics, marking a seminal milestone in the quantitative study of infectious diseases. Since then, generations of researchers have refined and extended his framework, greatly advancing our understanding of vector–host interactions and informing effective control strategies across diverse epidemiological settings \cite{Smith2012}. Nevertheless, despite its profound historical and theoretical influence, the Ross–Macdonald model and its subsequent extensions are not fully applicable to other vector-borne diseases such as DF and CHIKF.  

Unlike malaria, which is generally thought to be characterized mainly by symptomatic infection, DF and CHIKF involve both symptomatic and asymptomatic infections, and both groups can contribute to virus transmission \cite{Bustos2019, Duong2015}. Studies indicate that approximately 75\%–80\% of dengue infections are asymptomatic \cite{Bhatt2013}, while for chikungunya, the rate ranges from 15\% to 50\% \cite{Bustos2019, Ribeiro2025}. Because these individuals lack clinical symptoms, they often escape timely diagnosis and isolation. Nevertheless, their viremia levels remain sufficient to infect mosquitoes biting them \cite{Bustos2019, Duong2015, Scott2015, Phatchani2013}, thereby sustaining a hidden human–mosquito transmission cycle. This underscores that while vector elimination and source reduction are essential, they are necessary but insufficient. Preventing mosquito bites among asymptomatic carriers is equally crucial.

To figure out the role of asymptomatic carriers in chikungunya spread,  in Section \ref{sec2}, we develop a transmission model incorporating asymptomatic infections and a screening rate $q$ to evaluate how different initial  stages, human-induced, mosquito-induced, and joint transmission, shape the trajectory of outbreaks. Section \ref{sec3} analyzes both the long-term and short-term dynamics of the CHIKF model. It establishes a global stability threshold governed by $R_0$ and, for the critical short-term perspective, derives approximations for key outbreak characteristics (e.g., the peak value and the peak time) using singular perturbation theory, bridging the gap between transient dynamics and long-term predictions. 
Section \ref{sec4} focuses on the effect of the screening rate $q$ on the first and second waves of the epidemic, identifies an optimal screening rate $q_{\text{crit}}$, and analyzes how the value of $q$ affects whether a second wave emerges after the mosquito population rebounds. Finally, Section \ref{DisCon} discusses the implications of screening asymptomatic infected individuals for the current control strategy and proposes extending the ``Nine-character'' strategy to a ``Twelve-character'' strategy.

\section{Model formulation}\label{sec2}
 We divide the human population into four compartments, which are susceptible individuals ($S_h$), asymptomatic infected individuals ($I_{ha}$), symptomatic infected individuals ($I_{hs}$), and recovered individuals ($R_h$). The total human population is given by 
$N_h = S_h + I_{ha} + I_{hs} + R_h$, which remains constant in the absence of births and deaths. The adult female mosquito population is divided into susceptible ($S_m$) and infected ($I_m$) mosquitoes, with the total population denoted by $N_m = S_m + I_m$. Susceptible individuals ($S_h$) become infected through bites from infected mosquitoes ($I_m$). The force of infection follows a standard incidence form, depending on the number of infected mosquitoes, the proportion of susceptible humans, and the transmission efficiency $\beta_{hm}$ from mosquitoes to humans.   Here $\beta_{hm}$ denotes the probability that a susceptible human becomes infected after a single bite from an infectious mosquito. This parameter $\beta_{hm}$ is often interpreted as the proportion of mosquito pools in which the virus is present in saliva \cite{Bal2022}.  Both {\it Aedes aegypti} and {\it Aedes albopictus} are well-known vectors of CHIKV, with dissemination and transmission rates ranging between 40\% to 100\% \cite{Bal2022,  Manica2017}, i.e., $\beta_{hm}\in [0.4, 1]$.  We will employ the following equations to characterize the dynamics of the human population \begin{equation}\label{HumanE}
\left\{
\begin{aligned}
\frac{dS_h}{dt} &= -\beta_{hm} \frac{I_m}{N_h} S_h, \\
\frac{dI_{ha}}{dt} &= (1 - p) \beta_{hm} \frac{I_m}{N_h} S_h - \gamma_a I_{ha}, \\
\frac{dI_{hs}}{dt} &= p \beta_{hm} \frac{I_m}{N_h} S_h - \gamma_s I_{hs}, \\
\frac{dR_h}{dt} &= \gamma_a I_{ha} + \gamma_s I_{hs}.
\end{aligned}
\right.
\end{equation}  
In model (\ref{HumanE}), we assume that once infected, a fraction $p$ of individuals develop symptoms and enter the $I_{hs}$ class, while the remaining fraction $1-p$ becomes asymptomatic and enters $I_{ha}$ class. Although some studies suggest that most infected people exhibit symptoms \cite{Scott2015}, others report much higher proportions of asymptomatic individuals, including 47.1\% in \cite{Phatchani2013} and over 50\% in \cite{Ribeiro2025}.   Infected individuals recover at rates $\gamma_a$ (asymptomatic) and $\gamma_s$ (symptomatic). Although some patients may experience prolonged symptoms, CHIKF is generally self-limiting, with a typical illness duration of 7–14 days \cite{Cag2013NewM, Constant2021, Kumar2021}.  Thus, we take $\gamma_a, \gamma_s \in [1/14, 1/7]$. 

 For the purpose of this study, we make a deliberate assumption that symptomatic humans are effectively isolated and do not contribute to further mosquito infections. This is an ideal scenario designed to isolate the role of asymptomatic carriers. We acknowledge that in reality, symptomatic individuals may also infect mosquitoes if isolation is incomplete. Our assumption represents a best-case baseline, and relaxing it will be considered in future extensions of the model. Therefore, only asymptomatic infected humans ($I_{ha}$) can transmit the virus to susceptible mosquitoes ($S_m$). The transmission efficiency from humans to mosquitoes is denoted by $\beta_{mh}$,  which represents the probability that a susceptible mosquito becomes infected after feeding on an infectious human. Experimental studies show that $\beta_{mh}$ varies widely from 3.6\% to 94.7\%, depending on temperature and viral dose \cite{Alto2017, Vega2015}. The mosquito population dynamics are given by
\begin{equation}\label{MosE}
\left\{
\begin{aligned}
\frac{dS_m}{dt} &= \Lambda_m - \beta_{mh} \frac{(1 - q) I_{ha}}{N_h} S_m - \mu_m S_m, \\
\frac{dI_m}{dt} &= \beta_{mh} \frac{(1 - q) I_{ha}}{N_h} S_m - \mu_m I_m, 
\end{aligned}\right.
\end{equation}
where we assume that mosquitoes are recruited at a constant rate $\Lambda_m$ and die at a natural mortality rate $\mu_m$. The adult mosquito lifespan ranges from 7 to 30 days, so $\mu_m \in [1/30, 1/7]$ \cite{Blanco-Sierra2023, Cui2021, Yang2020}. The parameter $q \in [0,1]$ represents the rate at which asymptomatic individuals are identified and isolated through screening.  The factor $(1-q)$ in the transmission term reflects that only unscreened asymptomatic carriers remain capable of infecting mosquitoes. A higher $q$ means more effective detection and reduced transmission from asymptomatic cases. Although the U.S. CDC estimates that  15\%–35\%  of CHIKV infections are asymptomatic \cite{CDCchikungunya}, this may not reflect the situation in Guangdong, where limited screening could underestimate the true proportion. In fact, the Guangdong CDC only began reporting asymptomatic infections in October 2025 \cite{ChinanewsChikungunya}, nearly three months after the first imported case. Since asymptomatic individuals can still infect mosquitoes, random testing among asymptomatic and susceptible groups is a crucial public health measure.

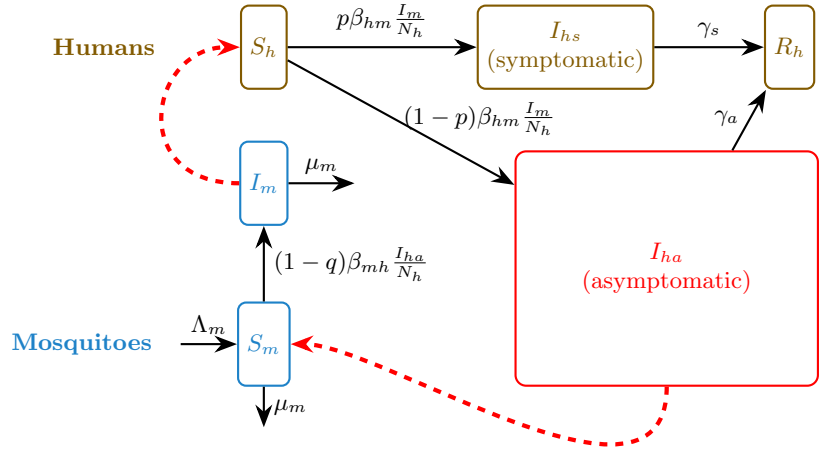
\begin{figure}[htbp]
\hspace{-2cm}  
\centering
\begin{tikzpicture}[
    node distance=2.6cm and 2.4cm,
    compartment/.style={rectangle, draw, rounded corners=3pt, thick,
        minimum height=1.1cm, align=center},
    bigcompartment/.style={rectangle, draw, rounded corners=4pt, thick,
        minimum width=4cm, minimum height=3.1cm, align=center},
    arrow/.style={-{Stealth[length=3mm]}, thick},
    dashedarrow/.style={-{Stealth[length=3mm]}, dashed, red!80!black, thick},
    font=\small
]

\definecolor{human}{RGB}{130,90,0}
\definecolor{asymptomatic}{RGB}{255,0,0}
\definecolor{mosquito}{RGB}{30,130,200}

\node[compartment, draw=human, text=human] (Sh) {$S_h$};
\node[compartment, draw=human, text=human, right=0cm and 2.5cm of Sh] (Ihs) {$I_{hs}$\\(symptomatic)};
\node[bigcompartment, draw=asymptomatic, text=asymptomatic, below right=0.8cm and 3cm of Sh] (Iha) {$I_{ha}$\\(asymptomatic)};
\node[compartment, draw=human, text=human, right=6.3cm of Sh] (Rh) {$R_h$};

\node[compartment, draw=mosquito, text=mosquito, below=2.8cm of Sh] (Sm) {$S_m$};
\node[compartment, draw=mosquito, text=mosquito, above=1cm of Sm] (Im) {$I_m$};

\draw[arrow] (Sh) -- node[above] {$p\beta_{hm}\tfrac{I_m}{N_h}$} (Ihs);
\draw[arrow] (Sh) -- node[above right=-3mm and -1mm] {$(1-p)\beta_{hm}\tfrac{I_m}{N_h}$} (Iha);
\draw[arrow] (Ihs) -- node[above] {$\gamma_s$} (Rh);
\draw[arrow] (Iha) -- node[above left=-2mm and 0mm] {$\gamma_a$} (Rh);

\draw[arrow] (Sm) -- node[right] {$(1-q)\beta_{mh}\tfrac{I_{ha}}{N_h}$} (Im);
\draw[arrow] (Sm) -- ++(0,-1.1) node[midway,right] {$\mu_m$};
\draw[arrow] (Im) -- ++(1.2,0) node[midway,above] {$\mu_m$};
\draw[arrow] ($(Sm)+(-1.1,0)$) -- node[midway,above] {$\Lambda_m$} (Sm);

\draw[dashedarrow, ultra thick, red, looseness=1.3] (Im.west) .. controls +(left:1.5cm) and +(left:1.2cm) .. (Sh.west);
\draw[dashedarrow, ultra thick, red, looseness=1.3] (Iha.south) .. controls +(down:2cm) and +(right:1.2cm) .. (Sm.east); 


\node[left=1cm of Sh, text=human] {\textbf{Humans}};
\node[left=1cm of Sm, text=mosquito] {\textbf{Mosquitoes}};

\end{tikzpicture}
\vspace{-0.2cm} 
\caption{{\bf Model structure with $q$ screening among asymptomatic/susceptible humans.} }
\label{FigSche}
\end{figure}

Model (\ref{HumanE})-(\ref{MosE})  can generate different transmission dynamics of CHIKF under different initial conditions at $t=0$, which are denoted by 
$$\mathbf{X}(0) = \big(S_h(0),\, I_{ha}(0),\, I_{hs}(0),\, R_h(0),\, S_m(0),\, I_m(0)\big),$$ 
with $N_h(0)=S_h(0)+I_{ha}(0)+I_{hs}(0)+R_h(0)$ and $N_m(0)=S_m(0)+I_m(0)$.   We will employ model (\ref{HumanE})-(\ref{MosE}) to analyze the dynamics of CHIKF under the national strategy and assess the risk posed by asymptomatic infected individuals. See Figure \ref{FigSche} for an illustration.  Table~\ref{Tab-Par} summarizes all epidemiological and entomological parameters used in model (\ref{HumanE})-(\ref{MosE}), together with their respective ranges and supporting references.

\begin{table}[h]
\centering
\caption{{\bf Epidemiological and entomological parameters in model (\ref{HumanE})-(\ref{MosE}) .}}
\label{Tab-Par}
\begin{tabularx}{\textwidth}{c>{\centering\arraybackslash}Xcc}
\toprule
\textbf{Parameter} & \textbf{Description (Unit)} & \textbf{Range} & \textbf{References} \\
\midrule
$\beta_{hm}$ & Transmission efficiency from infected mosquito to human & [0.4, 1.0] & \cite{Bal2022, Manica2017} \\
$p$ & Proportion of symptomatic cases among infected humans & [0.50, 0.85] & \cite{Scott2015, Phatchani2013, Ribeiro2025} \\
$\gamma_a$ & Recovery rate for asymptomatic individuals (day$^{-1}$) & [1/14, 1/7] & \cite{Cag2013NewM, Constant2021, Kumar2021} \\
$\gamma_s$ & Recovery rate for symptomatic individuals (day$^{-1}$)  & [1/14, 1/7] & \cite{Cag2013NewM, Constant2021, Kumar2021} \\
$\Lambda_m$ & Mosquito recruitment rate  (mosquito $\cdot$ day$^{-1}$)  & Adjustable & -- \\
$q$ & Screening rate among asymptomatic and susceptible individuals & Adjustable & -- \\
$\beta_{mh}$ & Transmission efficiency from infected human to mosquito & [0.14, 0.85] & \cite{Alto2017, Vega2015} \\
$\mu_m$ & Mortality rate of adult mosquitoes  (day$^{-1}$)  & [1/30, 1/7] & \cite{Blanco-Sierra2023, Cui2021, Yang2020} \\
\bottomrule
\end{tabularx}
\end{table}
 

Typically, a chikungunya outbreak is characterized by a sudden and significant increase in cases, far exceeding normal levels, as the virus spreads locally from person to person through the bites of mosquitoes \cite{LinkofCDCOutbreakInvestigation, LinkofHPSCCaseDefinitions}.  An epidemic outbreak typically progresses through three stages: onset, development, and eventual subsidence. To characterize the dynamics of an outbreak from different perspectives, relevant indicators include the epidemic peak, the time to peak, the final epidemic size, and the duration of an epidemic wave. This study focuses on the theoretical characterization of the epidemic peak and the time to peak.  Dynamics of model  (\ref{HumanE})-(\ref{MosE})  are highly dependent on how the infection is initially introduced. Based on the primary source of infection, we can categorize the initial conditions into three distinct types, each driven by a different combination of infectious humans and mosquitoes.

\textbf{(1) Human-induced transmission} ($I_{ha}(0)>0$ and $I_m(0)\approx0$).  Under this  stage, the outbreak begins with $I_{ha}$ but almost no established infected mosquito population. The initial growth may be slow and requires mosquitoes to act as a bridge population. The asymptomatic individuals $I_{ha}$ serve as the sole reservoir for seeding the mosquito population ($I_m$). A delayed peak may characterize the epidemic trajectory, as the system needs time to build up a sufficient force of infection within the mosquito population before a large-scale epidemic can occur in the human population. If human-to-mosquito transmission is inefficient (e.g., high $q$), the outbreak may fail to occur.

 \textbf{(2) Mosquito-induced transmission} ($I_m(0)>0$ and $I_{ha}(0)\approx0$).
In this case, the presence of infected mosquitoes $I_m(0) > 0$ provides an immediate source of infection for the susceptible human population $S_h$. This may lead to a rapid increase in human cases ($I_{ha}$ and $I_{hs}$) from the very beginning. The outbreak peaks earlier and with greater intensity compared to the human-induced case because the transmission cycle bypasses the initial bottleneck of infecting mosquitoes. The dynamics are initially dominated by the force of infection from mosquitoes to humans. 

\textbf{(3) Joint human--mosquito transmission} ($I_{ha}(0)>0$ and $I_m(0)>0$).
The joint transmission  stage represents the most favorable conditions for an epidemic. With both infectious humans and mosquitoes present at the onset, the complete transmission cycle is already operational. This may result in the most rapid epidemic growth and the highest peak incidence. The feedback loop between $I_{ha}$ and the mosquito population is immediately reinforced, potentially leading to a synchronized, explosive outbreak. The system quickly reaches its maximum transmission potential, often resulting in the largest final epidemic size among the three  stages. 

In summary, the initial source of infection determines the outbreak pattern. Human-induced outbreaks start slowly with delayed and lower peaks. Mosquito-induced outbreaks begin rapidly and are more intense, and joint introductions lead to the fastest and most severe epidemics. Mosquito population size and the presence of asymptomatic infections are key factors driving the early spread and potential for multiple waves of infection. 
 
\section{Model dynamics}\label{sec3}


\subsection{Asymptotic (long-term) dynamics} 
Through the analysis of model~(\ref{HumanE})-(\ref{MosE})  in  \ref{AppA}, the model has a disease-free equilibrium (DFE), denoted by $E_0 = (N_h, 0, 0, 0, \Lambda_m/\mu_m, 0)$,  from which one can calculate the basic reproduction number $R_0$ in \eqref{R0}. 
Dynamics of model (\ref{HumanE})-(\ref{MosE})   can be characterized by the basic reproduction number \( R_0 \). When \( R_0 < 1 \), $E_0$ is globally asymptotically stable (GAS), and if \( R_0 > 1 \), $E_0$ becomes unstable. This finding carries significant public health implications. Regardless of the initial epidemic conditions or the specific control strategies employed, if \( R_0 < 1 \), then the epidemic will eventually be contained, the outbreak will gradually subside, and the disease may be eradicated from the population. Hence, an epidemic does not necessarily evolve into an endemic state. When $R_0 > 1$, there is a unique stable endemic equilibrium, meaning the disease will become endemic over time. 

The global asymptotic stability of the disease-free equilibrium $E_0$ implies that infections eventually die out when $R_0<1$. However, this long-term result alone does not fully characterize the epidemic dynamics over a finite time horizon. In practice, even when an epidemic eventually subsides, the number of infections may continue to increase for some time and may exhibit one or more epidemic peaks. From a public health perspective, several indicators are relevant for understanding such transient epidemic dynamics, including the epidemic peak, the time to peak, the final epidemic size, and the duration of an epidemic wave. The epidemic peak and the time to peak characterize the maximum instantaneous infection burden and the time at which it occurs, respectively, whereas the final epidemic size and the duration of an epidemic wave describe the cumulative burden and the temporal extent of an outbreak, respectively. For the classical susceptible--infectious--recovered (SIR) model, the monograph by Brauer and Feng \cite{Brauer2019} and related studies \cite{Bai2023, Gimenez-Mujica2023} provide methods for estimating epidemic outcomes. However, explicit analytical characterization is more challenging for the high-dimensional vector--host (human--mosquito) coupled model considered in this study, because of the interactions among human--mosquito transmission, asymptomatic infections, and the screening mechanism. Accordingly, the subsequent analysis focuses on the theoretical characterization of the epidemic peak and the time to peak.


\subsection{Short-term (transient) dynamics } 

In general, the transient dynamics of dynamical systems are rather complex and, in the presence of periodic solutions, are seldom researched. However, in mosquito-borne disease models with a globally stable equilibrium, transient dynamics exhibit finite-time behavior and can approximate asymptotic dynamics. That is, the transient dynamics can be used to approximate the asymptotic dynamics. Under this framework, the transient dynamics of a mosquito-borne disease model can undergo, at most, a finite number of outbreaks. 
This section focuses on the approximate expressions for the epidemic peak and peak time when $q$ equals or is sufficiently close to 1.


  \ref{AppB2} considers the extreme case of complete screening ($q=1$). Under this condition, model (\ref{HumanE})-(\ref{MosE}) can be approximated by the single-population model (\ref{B2}), from which an explicit expression for the peak time can be derived: 
\begin{equation}\label{tp1} 
t_p^{(0)} = \frac{1}{\mu_m} \ln \frac{\beta_{hm} I_m(0) (N_h - I_{h,p}^{(0)})}{[\gamma_a(1-p)+\gamma_sp] I_{h,p}^{(0)}N_h}.
\end{equation} 

We then consider the case where $q=1-\epsilon$ with $\epsilon>0$ being sufficiently small in \ref{AppB3}. Using the method of singular perturbation theory, we approximate the peak value and the peak time as
\begin{align} 
I_{h,p}(\epsilon) &= I_h^{(0)}(t_p^{(0)}) + \epsilon I_h^{(1)}(t_p^{(0)}) + O(\epsilon^2),  \label{final_Ip} \\
t_p(\epsilon) &= t_p^{(0)} - \epsilon \left. \dfrac{d I_h^{(1)}}{dt} \right|_{t=t_p^{(0)}} \left( \left. \dfrac{d^2 I_h^{(0)}}{dt^2} \right|_{t=t_p^{(0)}} \right)^{\!-1} + O(\epsilon^2). \label{final_tp}
\end{align}
These three expressions theoretically describe the impact of the screening rate on CHIKF transmission and control when $q$ is sufficiently close to 1. For the general case with $q\in (0, 1)$, we will perform numerical explorations of model (\ref{HumanE})-(\ref{MosE})  in the next section. 
 
\section{Impact of screening on the transmission of CHIKV}\label{sec4}  
It should be emphasized that the following simulations are intended to illustrate mechanisms and are not calibrated to the actual outbreak in Guangdong. Our purpose is twofold: first, to show that asymptomatic infected individuals can sustain transmission even under vector suppression; and second, to reveal the possible existence of a critical screening threshold; rather than to quantitatively reproduce the observed epidemic curve. 

\subsection{The effect of $q$ on the first wave of outbreaks}\label{sec41} 
We focus on the first wave of an epidemic due to its unique epidemiological significance and critical window for intervention. Rapid and robust interventions at this stage, including case identification, isolation, transmission interruption, and public alerts, can effectively slow the spread, flatten the peak incidence, reduce the overall number of infections, and buy crucial time for vaccine development, medical resource allocation, and the design of more targeted control strategies. From both cost-effectiveness and public health protection perspectives, controlling the first wave is far more impactful than managing subsequent, potentially more complex waves. In this section, we investigate how the screening rate $q$ among asymptomatic and susceptible individuals influences both the timing and magnitude of the first epidemic wave’s peak.
 
To the end, we initiate model (\ref{HumanE})-(\ref{MosE})  with  
\begin{equation}\label{ini3} 
S_h(0)=995, \ I_{ha}(0)=5, \ I_{hs}(0)=0, \ R_h(0)=0, \ S_m(0)=480, \ I_m(0)=20,   
\end{equation} 
and parameters are taken as 
\begin{equation}\label{para3} 
\beta_{hm}=0.7, \ \beta_{mh}=0.5, \ p=0.7, \ \gamma_a=1/10,\ \gamma_s=1/7, \ \mu_m=1/10, \ \Lambda_m=50. 
\end{equation}  

 The initial conditions specified in (\ref{ini3}) represent an early epidemic  stage. That is, a largely susceptible human population ($>99\%$ susceptible) with minimal initial infections (only 0.5\% initially infected), and a mosquito population at its carrying capacity $K_m = \Lambda_m/\mu_m = 500$, with only 4\% of mosquitoes initially infected. These values are chosen to illustrate the dynamics during the most sensitive phase of epidemic emergence. The combination of high transmission rates ($\beta_{hm} = 0.7$, $\beta_{mh} = 0.5$), significant symptomatic proportion ($p = 0.7$), and biologically plausible recovery rates ($\gamma_a = 1/10$, $\gamma_s = 1/7$) creates an epidemic system where the asymptomatic reservoir, modulated by $q$, plays a determinative role in first wave progression. 
 
 To investigate whether asymptomatic infections can sustain disease transmission under strong vector control, we numerically solved model (\ref{HumanE})--(\ref{MosE}) using the initial conditions and parameter values specified in (\ref{ini3}) and (\ref{para3}), with the only modification being a reduction in the mosquito recruitment rate from $\Lambda_m=50$ to $\Lambda_m=5$. Since the demographic equilibrium of the mosquito population is given by $N_m^*=\Lambda_m/\mu_m$, this tenfold reduction in recruitment decreases the equilibrium mosquito population from $500$ to $50$, with the mosquito mortality rate held constant. Thus, reducing $\Lambda_m$ represents a substantial reduction in mosquito recruitment and, consequently, in the potential vector population, rather than the complete elimination of mosquitoes. Under this scenario, with $q=0$, the total number of infected individuals, $I_{ha}+I_{hs}$, reaches a peak of approximately $366$ on day $17.69$. Meanwhile, the total mosquito population decreases from its initial value of $500$ to approximately $127$, representing a reduction of about $74.6\%$ relative to the initial mosquito population. These results indicate that, despite the substantial decline in the mosquito population, the epidemic continues to develop. Therefore, substantial but incomplete vector control may not be sufficient to prevent the continued transmission of the disease over the short term, as long as asymptomatic individuals can still transmit the pathogen to mosquitoes.

Furthermore, by numerically solving model (\ref{B2}) with initial conditions $I_{ha}(0)=5$, $I_{hs}(0)=0$, and $I_m(0)=20$, the true peak value and time of total human infections are obtained as $I_{h,p}^{(0)}=45.58$ and $t_p^{(0)}=7.83$ days (see \ref{AppB}). The theoretical estimate of $t_p^{(0)}$ is $8.13$ days, yielding a relative error of $3.83\%$. 
For $q=0.95$, solving model (\ref{B5}) with $\epsilon=0.05$, $N_h=1000$, and $N_m=500$ gives a true peak $I_{h,p}=47.68$ and $t_p=9.08$ days. Using the zeroth-order solution $(I_{ha}^{(0)}, I_{hs}^{(0)}, I_m^{(0)})$ and solving the first-order correction model (\ref{1R}) with zero initial perturbations, we obtain $I_{h,p}^{(1)}=140.59$ and $t_p^{(1)}=28.03$ days. Thus, the first-order approximations give 
$
I_{h,p}^{(0)}+\epsilon I_{h,p}^{(1)}=52.61\ \text{(relative error: 10.30\%)}, \quad 
t_p^{(0)}+\epsilon t_p^{(1)}=9.23\ \text{days (relative error: 1.65\%)},
$
which shows that the first-order perturbation captures the peak time well but overestimates the peak magnitude.

To improve accuracy, a first-order correction is introduced in \ref{AppB3}, leading to equations (\ref{final_tp}) and (\ref{final_Ip}). By calculating
$$\left.\cfrac{dI_h^{(1)}}{dt}\right|_{t=t_p^{(0)}} = 7.98,\quad \left.\cfrac{d^2I_h^{(0)}}{dt^2}\right|_{t=t_p^{(0)}} =-0.58,\quad I_h^{(0)}(t_p^{(0)}) = 45.65,\quad I_h^{(1)}(t_p^{(0)}) = 37.19, $$ 
we obtain $t_p(\epsilon)=9.03$ days (relative error: $0.55\%$) and $I_{h,p}(\epsilon)=47.51$ (relative error: $0.36\%$). These results confirm that the corrected perturbation approach substantially improves accuracy, reducing relative errors to below 1\%.
 
\begin{figure}[htbp]
 \centering
  \includegraphics[height=0.3\textheight]{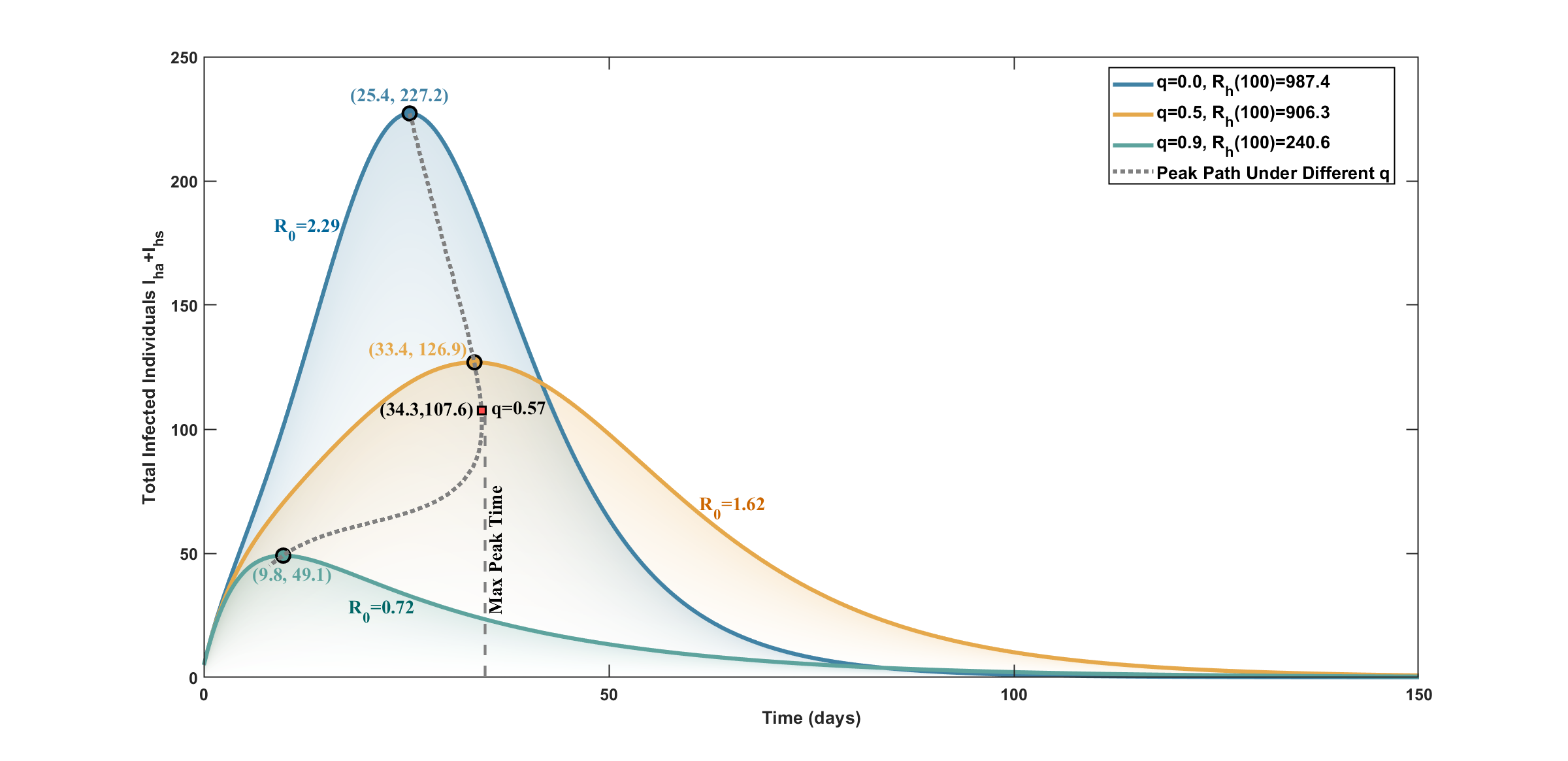}  
\caption{{\bf Effect of the screening rate ($q$) on the peak value and the peak time of the first epidemic wave.} Temporal dynamics of the total human infections ($I_h = I_{ha} + I_{hs}$) for different screening rates ($q = 0, 0.5, 0.9$), with the corresponding peak times and peak values indicated. The numerical simulations suggest that though $q=0.9$ or higher $q$ will end the outbreak sooner, there is an optimal value of $q=0.57$ when considering the social and medical resources needed for screening.}
\label{Fig3} 
\end{figure}

We now turn to examine the broader epidemiological patterns revealed by our model simulations. Figure \ref{Fig3} shows the human infection dynamics for different values of $q$. Specifically, when $q=0$, a simple calculation yields $R_0=2.29>1$, indicating that the transmission potential is high and leading to a rapid outbreak. The epidemic peaks of $I_h(t)$ in less than a month ($t_{p}=25.4$ days), infecting approximately 99\% of the population (the final size, which is defined as $R_h(100)= 986.8$ when $q=0$), with a peak number of infections reaching $I_{h, p}=227.2$, exhibiting a typical rapid outbreak pattern. For $q=0.4$,  we still have $R_0=1.62>1$, but the peak time is delayed to $t_{p}=34.5$ days, the peak value decreases to $I_{h, p}=152.0$, and the final size is reduced to $935.7$. The case of $q=0.9$ leads to $R_0=0.72<1$. In this case, the peak time advances to $t_{p}=9.8$ days, with a peak value $I_{h, p}=49.1$, and the final size drastically reduces to $240.6$. This phenomenon indicates that high-intensity detection can rapidly identify and block asymptomatic infections, significantly reducing the density of both human and mosquito infection sources and effectively cutting off the transmission chain.

These numerical observations reveal the complex, nonlinear effects of the screening rate on mosquito-borne diseases. The screening rate does not follow a monotonic pattern in which higher values always yield better outcomes. Instead, there exists an optimal critical value $q_{\text{crit}}\approx 0.57$. When resources are limited, targeting  $q_{\text{crit}}$ can achieve maximum epidemic delay, buying valuable time for containment efforts. When resources are sufficient, adopting a high screening rate can rapidly control the outbreak and reduce the peak magnitude. This non-monotonic relationship underscores the importance of precision in disease containment strategies. It provides a scientific basis for public health decision-making by showing that different containment goals require different screening strategies. Delaying the epidemic requires a critical screening rate, whereas controlling the outbreak necessitates a high screening rate.  

 We should mention that the specific value $q_{crit}=0.57$ is obtained for the chosen parameter set and should not be interpreted as a universal threshold. Rather, it serves to demonstrate that such a critical value can exist, and that the relationship between screening effort and epidemic delay is non-monotonic, a qualitative insight that is robust across reasonable parameter ranges. 
\subsection{The effect of $q$ on the second wave of outbreaks}\label{Result3} 
 Based on the analysis of the first wave, we now examine how the screening rate $q$ affects the emergence and severity of a second wave. We define the ``second wave'' quantitatively as a renewed rise in daily new infections after the first wave subsides, forming a distinct secondary peak. Classical epidemiological theory holds that if the first wave is large enough, herd immunity can be established, thereby preventing a severe second wave \cite{Smith2012}. Conversely, if the first wave is too small and a large susceptible population remains when transmission slows, then once the mosquito population rebounds (e.g., due to relaxed control or climatic conditions favoring mosquito breeding), and given insufficient screening (i.e., $q$ too low), a second wave more severe than the first may occur. Traditional approaches often emphasize intensive surveillance and high screening rates, but these measures are resource-intensive and difficult to sustain over the long term. This subsection examines whether a moderate screening rate can effectively prevent a second wave, thereby providing a more sustainable strategy for long-term control. To this end, we set the initial conditions of model (\ref{HumanE})-(\ref{MosE}) as follows: 
\begin{equation}\label{ini2} 
S_h(0)=950, \ I_{ha}(0)=20, \ I_{hs}(0)=30, \ R_h(0)=0, \ S_m(0)=400, \ I_m(0)=50,   
\end{equation} 
and the parameters are taken as 
\begin{equation}\label{para2}
\beta_{hm}=0.4, \  \beta_{mh}=0.35, \  p=0.6, \  \gamma_a=1/13,\  \gamma_s=1/7, \  \mu_m=1/10, \  \Lambda_m=15.
\end{equation} 
On day 60, a fourfold increase in the mosquito recruitment rate is implemented to simulate a resurgence of the mosquito vector population.   Here, we artificially increase mosquito recruitment to illustrate a plausible scenario in which vector control measures are relaxed or environmental conditions become more favorable for mosquito breeding. The exact timing is chosen for illustration only, and the key point is that such a rebound can occur if asymptomatic carriers remain undetected. 

\begin{figure}[htbp]
 \centering
  \includegraphics[height=0.3\textheight]{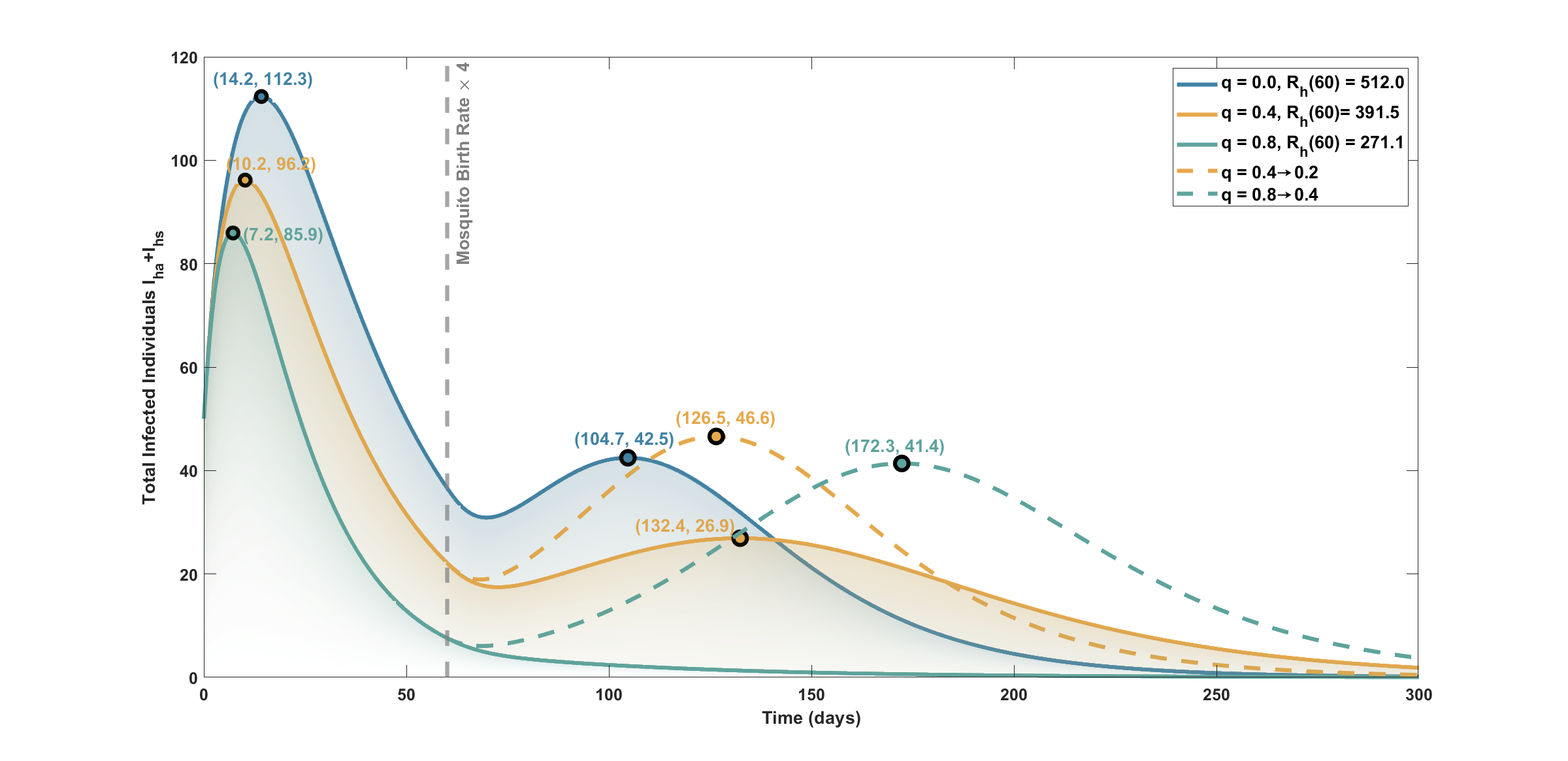}  
\caption{{\bf Effect of the screening rate ($q$) on the second epidemic wave.} Model (\ref{HumanE})-(\ref{MosE})  is initiated with (\ref{ini2}) and parameters are given in (\ref{para2}). Temporal dynamics of the total human infections ($I_h = I_{ha} + I_{hs}$) for different screening rates ($q = 0, 0.4, 0.8$), with the corresponding peak times and peak values indicated. The mosquito population is fourfold boosted at day 60 (indicated by the vertical dashed line), and the pair of dashed lines indicates the shift to a lower screening strategy at $t=60$.  }
\label{FigBo} 
\end{figure} 

The simulation results, shown in Figure \ref{FigBo}, reveal a compelling pattern. At a screening rate of $q = 0.4$, the second wave peak measures only 26.9 individuals, representing just 28$\%$ of the first wave's magnitude (96.2 individuals). More importantly, when we strategically reduce screening from $q = 0.4$ to $q = 0.2$ after the first wave, the second wave peak increases to 46.6 individuals, yet this still represents effective control as it remains well below the first wave peak and demonstrates that moderate screening reduction during the second wave does not lead to catastrophic subsequent outbreaks. This phenomenon is explained by the susceptible population dynamics. At $q = 0.4$, the susceptible population ($S_h$) measures 586.3 individuals at day 60, compared to 391.5 recovered individuals ($R_h$). The immunity acquired during the first wave reduced the number of susceptible hosts available in the second wave, thereby limiting the scale of transmission. 

Furthermore, our numerical trials clearly demonstrate that excessively high screening rates are counterproductive for long-term epidemic control. While $q = 0.8$ screening eliminates the second wave (no detectable peak), it comes at a substantial cost. That is, only 271.1 individuals recover by day 60, leaving 721.3 individuals still susceptible. This creates a large susceptibility pool in which the population remains highly vulnerable to future outbreaks if screening is relaxed, as evidenced by the substantial second-wave peak of 41.4 individuals when screening is reduced from $q = 0.8$ to $q = 0.4$ at $t = 60$. In contrast, the $q = 0.4$ strategy achieves a balanced outcome. It controls the first wave effectively (peak of 96.2 vs 112.3 at $q = 0$), maintains a manageable second wave (26.9 vs 42.5 at $q = 0$), and builds substantial population immunity (391.5 recovered individuals).  To compare the relative severity of resurgent outbreaks across different screening rates, we use the peak ratio, defined as the second wave peak divided by the first wave peak. This normalized measure removes the influence of varying first wave magnitudes and allows a fair comparison of how well each screening strategy controls resurgence relative to its initial epidemic impact. The peak ratio data provide further confirmation of this advantage, with a value of $q = 0.4$ yielding a peak ratio of 0.28 (26.9/96.2), which is significantly better than the ratio of 0.38 (42.5/112.3) observed at $q = 0$.

From a public health perspective, these findings support implementing moderate screening rates ($q = 0.4$) during the initial epidemic waves, followed by a strategic reduction to lower levels ($q = 0.2$) in subsequent waves. This approach delivers two key benefits. It effectively controls the immediate outbreak while simultaneously building sufficient population immunity to naturally mitigate future waves. For public health systems with limited resources, this strategy supports the optimal allocation of screening resources. It allows maintaining moderate screening intensity during the critical initial phases and reallocating those resources to other health priorities once substantial population immunity has been established.

\section{From the {\it Nine-character  Strategy} to the {\it Twelve-character Strategy}: Discussion and Conclusions}\label{DisCon} 
Motivated by the persistent disconnection between mosquito reduction efforts and the continued growth of human cases observed in the field, we developed a transmission model (\ref{HumanE})-(\ref{MosE})  that explicitly incorporates asymptomatic individuals and a screening rate. Model (\ref{HumanE})-(\ref{MosE}) demonstrates that the initial source of infection, whether from humans, mosquitoes, or both, profoundly influences the speed, scale, and trajectory of a chikungunya outbreak. Asymptomatic individuals play a central role in sustaining transmission, particularly when mosquito control is intensified. The screening rate  $q$ emerges as a key leverage point for interrupting hidden transmission chains.  These insights suggest that incorporating screening of asymptomatic cases may strengthen the current Nine-character strategy. We therefore propose, as a working hypothesis, an updated  {\it Twelve-character strategy}  for outbreak management. 
 
 Our analysis in Section \ref{sec3} and Section \ref{sec4} demonstrates that while the basic reproduction number $R_0$ determines the long-term fate of the epidemic, the short-term outbreak trajectory is highly sensitive to the screening rate $q$. The derived approximations for the peak value and the peak time underpin that proactive screening of asymptomatic infections is not merely beneficial but essential for mitigating outbreak intensity and delaying epidemic peaks, providing a mathematical basis for targeted public health intervention. 
 
 Numerical investigation in subsection \ref{sec41} reveals a sophisticated dual role of the asymptomatic screening rate $q$ in shaping the dynamics of the first epidemic wave, exhibiting marked nonlinear effects on both the peak timing and magnitude. Rather than following a simple linear relationship, $q$ displays a characteristic inverted U-shape influence on the peak time: below the critical threshold $q_{\text{crit}} \approx 0.57$, increasing detection delays the epidemic peak by reducing transmission efficiency without fully interrupting transmission chains, thereby providing crucial time for public health preparedness. Conversely, beyond $q_{\text{crit}}$, efficient detection rapidly identifies and isolates asymptomatic cases, leading to earlier epidemic suppression.  

 Our findings in subsection \ref{Result3} reveal that moderate screening rates ($q=0.4$) offer an optimal balance between immediate epidemic control and long-term population immunity. Unlike high screening rates that leave large susceptible pools, or low rates that cause severe initial outbreaks, the moderate approach effectively controls the first wave while building sufficient immunity to suppress subsequent waves naturally. The strategy of starting with moderate screening ($q=0.4$) and strategically reducing it ($q=0.2$) after the first wave proves particularly effective. It maintains outbreak control while allowing efficient reallocation of limited public health resources, providing a practical and sustainable solution for managing vector-borne disease outbreaks in resource-constrained settings.  This approach to dynamically allocating limited resources aligns with the principles of optimal control under constraints, as demonstrated in models of dengue management \cite{Ahmed2016}.  

In conclusion, this study provides a mathematical framework for understanding the complex interactions among vector control, asymptomatic transmission, and active screening in the dynamics of vector-borne diseases. The results indicate that traditional approaches relying solely on vector control have limitations, and that an integrated strategy is needed to intervene simultaneously in both the vector and human components of the transmission cycle. Under the parameter settings adopted here, if only vector-targeted measures are implemented, the epidemic scale may continue to grow for several weeks before peaking. However, our analysis reveals a possible alternative: rapidly intensified screening to identify asymptomatic infected individuals, for example by providing polymerase chain reaction or rapid antigen testing to high-risk groups \cite{PDHC}, combined with measures to prevent mosquito bites, may alter the epidemic trajectory, bring the epidemic into decline earlier, and reduce the final infection scale. Accordingly, this study extends the current ``Nine-character'' strategy to a ``Twelve-character'' strategy of ``clearing stagnant water, killing adult mosquitoes, preventing bites, and screening cases early,'' with the aim of providing a reference for the control of CHIK and other vector-borne diseases. Future research should incorporate spatial heterogeneity, human mobility, and economic factors to develop more realistic control strategies.

 \Acknowledgements{This work was supported by the National Natural Science Foundation of China (Nos: 12371484, 12331017, 12401633), GAST Youth Science and Technology Talent Cultivation Program (No. SKXRC2026676), China Postdoctoral Science Foundation (No. 2026M793353) and the Postdoctoral Fellowship Program of CPSF (No. GZC20261665). We thank the anonymous reviewers for their valuable comments and suggestions on this paper.} 


\begin{appendix} 
\section{Derivation of basic reproduction number $R_0$}\label{AppA} 
 The disease-free equilibrium (DFE) of model (\ref{HumanE})-(\ref{MosE})  is $E_0 = \left(N_h, 0, 0, 0,  \Lambda_m/\mu_m, 0\right). $ By using the next generation matrix method \cite{Asamoah2021}, we can directly calculate the basic reproduction number as
\begin{equation}\label{R0} 
R_0 = \sqrt{\frac{\beta_{hm}\beta_{mh}(1-q)(1-p)N_m}{\gamma_a\mu_m N_h}}. 
\end{equation} 
Furthermore, the value of $R_0$ can be decomposed into two distinct transmission components. The first component, denoted by $R_{hm}$, estimates the human-to-mosquito transmission efficiency factor, and
\begin{equation}\label{Rhm} 
R_{hm} = \sqrt{\frac{\beta_{mh}(1-q)}{\mu_m}}.
\end{equation} 
The second part estimated the mosquito-to-human transmission pressure factor, defined by 
\begin{equation}\label{Rmh} 
R_{mh} = \sqrt{\frac{\beta_{hm}(1-p)N_m}{\gamma_a N_h}}.
\end{equation}   
 Clearly, we have $R_0=R_{hm}R_{mh}$, where $R_{hm}$ captures human to mosquito amplification; and $R_{mh}$ captures vector mosquito to human pressure, and screening rate $q$ reduces  $R_{hm}$ multiplicatively. 
 
  \section{Theoretical characterization of the first epidemic peak} \label{AppB} 
 \subsection{The reduced system of  (\ref{HumanE})-(\ref{MosE}) }  \label{AppB1} 
Assuming rapid dynamics or focusing on the early epidemic phase where recovered individuals are negligible ($R_h \approx 0$), and substituting $S_h = N_h - I_{ha} - I_{hs}$ and $S_m = N_m - I_m$ into the infection equations of model (\ref{HumanE})-(\ref{MosE}), we obtain 
 \begin{equation}\label{B1} 
 \begin{cases}
\cfrac{dI_{ha}}{dt} &= (1-p)\beta_{hm}\cfrac{I_m}{N_h}(N_h - I_{ha} - I_{hs}) - \gamma_a I_{ha}, \\
\cfrac{dI_{hs}}{dt} &= p\beta_{hm}\cfrac{I_m}{N_h}(N_h - I_{ha} - I_{hs}) - \gamma_s I_{hs}, \\
\cfrac{dI_m}{dt} &= \beta_{mh}\cfrac{(1-q)I_{ha}}{N_h}(N_m - I_m) - \mu_m I_m. 
\end{cases}
\end{equation} 
 
\subsection{Peak magnitude and timing analysis of model (\ref{B1}) when $q=1$}\label{AppB2}
When $q=1$, model (\ref{B1}) is reduced to
 \begin{equation}\label{B2} 
\begin{cases}
\cfrac{dI_{ha}}{dt} = (1 - p)\beta_{hm}\cfrac{I_m}{N_h}(N_h - I_h) - \gamma_a I_{ha}, \\
\cfrac{dI_{hs}}{dt} = p\beta_{hm}\cfrac{I_m}{N_h}(N_h - I_h) - \gamma_s I_{hs}, 
\end{cases}
\end{equation}
in which $I_m$ satisfies  
$I_m(t) = I_m(0)e^{-\mu_m t}$.  Let $I_h= I_{ha} + I_{hs}$. We have  
\begin{equation}
\label{Ih_eq}
\cfrac{dI_h}{dt} = \beta_{hm}\cfrac{I_m}{N_h}(N_h - I_h) - (\gamma_a I_{ha} + \gamma_s I_{hs}).
\end{equation}
It is easy to see that the basic reproduction number governed by (\ref{Ih_eq}) is approximately 
$$R_e(t)=\frac{\beta_{hm}I_m(0)e^{-\mu_m t}}{\bar{\gamma}N_h}, \ \text{with}\ \bar{\gamma}=\frac{\gamma_a+\gamma_s}{2}. $$ 
Therefore, $R_e(0)>1$ is possible when $I_m(0)$ is large enough, together with a peak value of $I_h^{(0)}(t)  = I_{ha}^{(0)}(t) + I_{hs}^{(0)}(t)$, where $( I_{ha}^{(0)}(t), I_{hs}^{(0)}(t), I_m^{(0)}(t))$ is the solution of model (\ref{B2}), which satisfies $I_m^{(0)}(t)  = I_m(0)e^{-\mu_m t}$, and the human infections $I_{ha}^{(0)}(t)$ and $I_{hs}^{(0)}(t)$ need to be obtained by numerically solving the first two equations of (\ref{B2}). Let $I_{h, p}^{(0)}$ and $t_p^{(0)}$ be the peak value of $I_h^{(0)}$ and the time point hitting the peak value $I_{h, p}^{(0)}$, respectively. Then,  
\begin{equation}\label{B4} 
\beta_{hm} \frac{I_m{(0)} e^{-\mu_m t_p^{(0)}}}{N_h} (N_h - I_{h, p}^{(0)}) = \gamma_a I_{ha, p}^{(0)}+\gamma_s I_{hs, p}^{(0)}, 
\end{equation} 
given that  $I_{ha}$, $I_{hs}$ and $I_h$ hit the peak values simultaneously. 
It is reasonably good to assume that $I_{ha, p}^{(0)}=(1-p)I_{h, p}^{(0)}$ and $I_{hs, p}^{(0)}=pI_{h, p}^{(0)}$. Therefore, we get (\ref{tp1})  from (\ref{B4}).

%

\subsection{Peak magnitude and timing analysis of model (\ref{B1}) when $q\approx 1$}\label{AppB3}
Let $q=1-\epsilon$ with $\epsilon>0$ being small enough. In this case, model (\ref{B1}) becomes 
 \begin{equation}\label{B5} 
 \begin{cases}
\cfrac{dI_{ha}}{dt} &= (1-p)\beta_{hm}\cfrac{I_m}{N_h}(N_h - I_{ha} - I_{hs}) - \gamma_a I_{ha}, \\
\cfrac{dI_{hs}}{dt} &= p\beta_{hm}\cfrac{I_m}{N_h}(N_h - I_{ha} - I_{hs}) - \gamma_s I_{hs}, \\
\cfrac{dI_m}{dt} &= \beta_{mh}\cfrac{\epsilon I_{ha}}{N_h}(N_m - I_m) - \mu_m I_m. 
\end{cases}
\end{equation} 
Substituting $I_{ha}(t)  = I_{ha}^{(0)}(t) + \epsilon I_{ha}^{(1)}(t) +O(\epsilon^2)$, $I_{hs}(t)  = I_{hs}^{(0)}(t) + \epsilon I_{hs}^{(1)}(t) +O(\epsilon^2)$ and 
 $I_m(t)  = I_m^{(0)}(t) + \epsilon I_m^{(1)}(t) +O(\epsilon^2)$ into (\ref{B5}), we get    \begin{equation}\label{1R}
\begin{cases}
\cfrac{dI_{ha}^{(1)}}{dt} = (1-p)\beta_{hm}\left[\cfrac{I_m^{(1)}}{N_h}(N_h -  I_{ha}^{(0)}-I_{hs}^{(0)}) - \cfrac{I_m^{(0)}}{N_h}(I_{ha}^{(1)} + I_{hs}^{(1)})\right] - \gamma_a I_{ha}^{(1)}, \\
\cfrac{dI_{hs}^{(1)}}{dt} = p\beta_{hm}\left[\cfrac{I_m^{(1)}}{N_h}(N_h -  I_{ha}^{(0)}-I_{hs}^{(0)}) - \cfrac{I_m^{(0)}}{N_h}(I_{ha}^{(1)} + I_{hs}^{(1)})\right] - \gamma_s I_{hs}^{(1)}, \\
\cfrac{dI_m^{(1)}}{dt} = \beta_{mh}\cfrac{I_{ha}^{(0)}}{N_h}(N_m - I_m^{(0)}) - \mu_m I_m^{(1)}
\end{cases}
\end{equation}
to characterize the dynamics of  $I_{ha}^{(1)}(t), I_{hs}^{(1)}(t)$ and $I_m^{(1)}(t)$ with $I_{ha}^{(1)}(0)= I_{hs}^{(1)}(0)=I_m^{(1)}(0)$.  We assume that the total number of infected individuals $I_h(t) = I_{ha}(t) + I_{hs}(t)$ reaches its peak value $I_{h, p}(\epsilon)$ at time $t_p(\epsilon)$. Then,  
\begin{equation}\label{PC1}
\frac{dI_h}{dt}\bigg|_{t=t_p(\epsilon)} = 0, 
\end{equation}
and $I_{h, p}(\epsilon) = I_h(t_p(\epsilon))$. 
We expand $t_p(\epsilon)$ and $I_{h, p}(\epsilon)$ as power series in $\epsilon$:
\begin{align}
t_p(\epsilon) &= t_p^{(0)} + \epsilon t_p^{(1)} + O(\epsilon^2), 
\label{te} \\
I_{h, p}(\epsilon) &= I_{h, p}^{(0)}+ \epsilon I_{h, p}^{(1)} + O(\epsilon^2),  \label{Ie}
\end{align}
where $ I_{h, p}^{(0)} = I_h^{(0)}(t_p^{(0)})$. By expanding $dI_h/dt$ at $t=t_p(\epsilon)$, together with (\ref{PC1}), we  obtain 
\begin{equation}\label{t1_solution}
t_p^{(1)} = - \left. \dfrac{d I_h^{(1)}}{dt} \right|_{t=t_p^{(0)}} \left( \left. \dfrac{d^2 I_h^{(0)}}{dt^2} \right|_{t=t_p^{(0)}} \right)^{\!-1}.
\end{equation}

 For the peak magnitude $I_{h,p}(\epsilon)$, we have:
\begin{align*}
I_{h,p}(\epsilon) &= I_h(t_p(\epsilon))  = I_h^{(0)}(t_p(\epsilon)) + \epsilon I_h^{(1)}(t_p(\epsilon)) + O(\epsilon^2).  
\end{align*}

Expanding $I_h^{(0)}(t_p(\epsilon))$ around $t_p^{(0)}$:
\begin{align*}
I_h^{(0)}(t_p(\epsilon))   = I_{h,p}^{(0)} + \epsilon t_p^{(1)} \cdot \frac{dI_h^{(0)}}{dt}\bigg|_{t=t_p^{(0)}} + O(\epsilon^2). 
\end{align*}
Since at the peak point $t_p^{(0)}$ we have $\left.\cfrac{dI_h^{(0)}}{dt}\right|_{t=t_p^{(0)}} = 0$, it follows that:
$I_h^{(0)}(t_p(\epsilon)) = I_{h,p}^{(0)} + O(\epsilon^2)$. 
For the first-order term, we have $\epsilon I_h^{(1)}(t_p(\epsilon)) = \epsilon I_h^{(1)}(t_p^{(0)}) + O(\epsilon^2)$. 
Therefore, the peak magnitude is $I_{h,p}(\epsilon) = I_{h,p}^{(0)} + \epsilon I_h^{(1)}(t_p^{(0)}) + O(\epsilon^2)$. 
That is, $I_{h,p}^{(1)} = I_h^{(1)}(t_p^{(0)})$. Therefore, (\ref{final_tp}) and (\ref{final_Ip}) follow from (\ref{te}) and (\ref{Ie}), respectively. 

\end{appendix}

\end{document}